\documentclass[aps,prd,onecolumn,eqsecnum, nofootinbib]{revtex4-2}
\usepackage{amssymb}
\usepackage{amsmath}
\usepackage{amsfonts}
\usepackage{hyperref}
\DeclareSymbolFont{EulerScript}{U}{eus}{m}{n}
\SetSymbolFont{EulerScript}{bold}{U}{eus}{b}{n}
\DeclareSymbolFontAlphabet\scrpt{EulerScript}
\allowdisplaybreaks
\begin{document}
\title{Mukkamala-Pere\~niguez master function for even-parity perturbations of the Schwarzschild spacetime}  
\author{Eric Poisson}
\email{epoisson@uoguelph.ca}
\affiliation{Department of Physics, University of Guelph, Guelph, Ontario, N1G 2W1, Canada} 
\date{January 20, 2025}
\begin{abstract} 
Mukkamala and Pere\~niguez recently discovered a new master function for even-parity metric perturbations of the Schwarzschild spacetime. Remarkably, this function satisfies the Regge-Wheeler equation (instead of the Zerilli equation), which was previously understood to govern the odd-parity sector of the perturbation only. In this paper I follow up on their work. First, I identify a source term for their Regge-Wheeler equation, constructed from the perturbing energy-momentum tensor. Second, I relate the new master function to the radiation fields at future null infinity and the event horizon. Third, I reconstruct the metric perturbation from the new master function, in the Regge-Wheeler gauge. The main conclusion of this work is that the greater simplicity of the Regge-Wheeler equation (relative to the Zerilli equation) is offset by a greater complexity of obtaining the radiation fields and reconstructing the metric. 
\end{abstract} 
\maketitle

\section{Introduction} 
\label{sec:intro} 

Perturbations of the Schwarzschild spacetime were first considered in a seminal paper by Regge and Wheeler \cite{regge-wheeler:57}. In spite of the complexity of the linearized Einstein equations for these perturbations, Regge and Wheeler managed to decouple the equations that govern the odd-parity sector of the perturbation (also known as axial perturbations), and produced the famous Regge-Wheeler equation for a master function that encodes the odd-parity perturbations. They did not, however, decouple the equations for the even-parity sector of the perturbation (also known as polar perturbations). This was eventually achieved by Zerilli \cite{zerilli:70}, who encapsulated these perturbations within another master function, which satisfies a variant of the Regge-Wheeler equation known (appropriately) as the Zerilli equation. This even-parity function is now called the Zerilli-Moncrief function, after Moncrief established its gauge invariance \cite{moncrief:74}.  

In a striking recent development \cite{mukkamala-pereniguez:24}, Mukkamala and Pere\~niguez constructed a new master function for even-parity perturbations of the Schwarzschild spacetime, which satisfies the Regge-Wheeler equation instead of the Zerilli equation. This remarkable discovery, almost 55 years after Zerilli's own efforts, comes with an immediate conceptual benefit: the isospectrality of even-parity and odd-parity quasinormal modes of a Schwarzschild black hole is no longer to be viewed as a consequence of an intricate transformation between the Regge-Wheeler and Zerilli equations \cite{chandrasekhar-detweiler:75}, but as a trivial manifestation of the fact that even-parity and odd-parity perturbations are governed by the same master equation. There is also a practical benefit: the Regge-Wheeler potential is simpler than the Zerilli potential, and the increased simplicity of the master equation for even-parity perturbations of the Schwarzschild spacetime will facilitate analytical and numerical efforts to calculate these perturbations.  

In this paper I follow up on the remarkable discovery of Mukkamala and Pere\~niguez. First, I obtain the source term for their Regge-Wheeler equation, from the perturbing energy-momentum tensor. Second, I work out the relation between their master function and the radiation fields at future null infinity and the event horizon. Third, I consider the task of reconstructing the metric perturbation (in the Regge-Wheeler gauge) from their master function. My main conclusion is that while the Mukkamala-Pere\~niguez master function does satisfy a simpler differential equation than Zerilli's master function, this comes at the cost of a more convoluted relationship with the radiation fields and a more arduous procedure to reconstruct the metric perturbation. The advantages and disadvantages of each formulation (Mukkamala and Pere\~niguez versus Zerilli) depend on the context and the applications that are being considered.   

\section{Perturbation of the Schwarzschild spacetime}
\label{sec:background}

I adopt the notations and conventions of Martel and Poisson \cite{martel-poisson:05}. The perturbed metric is written as $g_{\alpha\beta} + p_{\alpha\beta}$, with $g_{\alpha\beta}$ denoting the background Schwarzschild metric and $p_{\alpha\beta}$ the perturbation. Arbitrary coordinates $x^a$ are used in the $t$-$r$ submanifold of the Schwarzschild spacetime, which comes with a metric $g_{ab}$; $g^{ab}$ is its inverse, and $\nabla_a$ is the covariant-derivative operator compatible with this metric. Angles $\theta^A = (\theta,\phi)$ are used in the orthogonal submanifold; $\Omega_{AB} := \mbox{diag}[1, \sin^2\theta]$ is the metric on the unit two-sphere, $\Omega^{AB}$ is its inverse, and $D_A$ is the covariant-derivative operator compatible with this metric. In this notation, the background Schwarzschild metric is written as
\begin{equation}
g_{\alpha\beta}\, dx^\alpha dx^\beta = g_{ab}\, dx^a dx^b + r^2\, \Omega_{AB}\, d\theta^A d\theta^B,
\end{equation}
where $g_{ab}$ and $r$ are functions of $x^a$. The timelike Killing vector of the spacetime is denoted $t^a$, and I set $r_a := \nabla_a r$; we have that $f := g^{ab} r_a r_b = 1 - 2M/r$. 

The even-parity sector of the metric perturbation is decomposed in spherical harmonics according to
\begin{subequations}
\label{decomp} 
\begin{align}
p_{ab} &= \sum_{\ell m} h^{\ell m}_{ab}\, Y^{\ell m}, \\
p_{aB} &= \sum_{\ell m} j^{\ell m}_a\, Y^{\ell m}_A, \\
p_{AB} &= r^2 \sum_{\ell m} \bigl( K^{\ell m}\, \Omega_{AB} Y^{\ell m}
+ G^{\ell m}\, Y^{\ell m}_{AB} \bigr),
\end{align}
\end{subequations}
in which the fields $h^{\ell m}_{ab}$, $j^{\ell m}_a$, $K^{\ell m}$, and $G^{\ell m}$ depend on the coordinates $x^a$ only. To unclutter the notation I shall henceforth omit the $\ell m$ label on these fields, and omit the summation sign in equations like Eq.~(\ref{decomp}). My considerations throughout this paper are limited to $\ell \geq 2$.

The spherical harmonics implicated in Eq.~(\ref{decomp}) are the familiar scalar harmonics $Y^{\ell m}(\theta,\phi)$, the vector harmonics
\begin{equation} 
Y^{\ell m}_A := D_A Y^{\ell m},
\end{equation}
and the tensor harmonics
\begin{equation}
Y^{\ell m}_{AB} := \bigl[ D_A D_B - \tfrac{1}{2} \ell(\ell+1) \Omega_{AB} \bigr] Y^{\ell m};
\end{equation}
these are tracefree by virtue of the eigenvalue equation satisfied by the scalar harmonics, $\Omega^{AB} Y^{\ell m}_{AB} = 0$. The scalar harmonics are taken to be normalized, and the normalization of the vector and tensor harmonics is described by Eqs.~(3.3) and (3.8) of Martel and Poisson.

The gauge freedom associated with metric perturbations allows us to set $j_a = 0 = G$; this defines the Regge-Wheeler gauge. The combinations
\begin{subequations}
\begin{align}
\tilde{h}_{ab} &= h_{ab} - \nabla_a \varepsilon_b - \nabla_b \varepsilon_a, \\
\tilde{K} &= K + \tfrac{1}{2} \ell(\ell+1) G - 2 r^{-1} r^a \varepsilon_a,
\end{align}
\end{subequations}
with $\varepsilon_a := j_a - \frac{1}{2} r^2 \nabla_a G$, are gauge invariant. They become $\tilde{h}_{ab} = h_{ab}$ and $\tilde{K} = K$ when the Regge-Wheeler gauge is adopted.

The (linearized) Einstein tensor constructed from the perturbed metric is decomposed in spherical harmonics according to
\begin{subequations}
\begin{align}
G^{ab} &= Q^{ab}\, Y^{\ell m}, \\
G^{aB} &= \frac{1}{2 r^2} Q^a\, \Omega^{BD} Y^{\ell m}_D, \\
G^{AB} &= \frac{1}{2 r^2} Q^\flat\, \Omega^{AB} Y^{\ell m}
+ \frac{1}{2 r^4} Q^\sharp\, \Omega^{AC} \Omega^{BD} Y^{\ell m}_{CD},
\end{align}
\end{subequations}
where I again omit the $\ell m$ labels and summation signs on the right-hand sides; the factors of $1/2$, $r^{-2}$ and $r^{-4}$ are inserted for convenience. The objects $Q^{ab}$, $Q^a$, $Q^\flat$, and $Q^\sharp$ are linear differential operators acting on the gauge-invariant fields $\tilde{h}_{ab}$ and $\tilde{K}$. Explicit expressions are given by Eqs.~(4.13)--(4.16) of Martel and Poisson. (Because the Einstein tensor vanishes in the background Schwarzszschild spacetime, its perturbation is necessarily gauge invariant.)  

The Einstein field equations imply that $Q^{ab}$, $Q^a$, $Q^\flat$, and $Q^\sharp$ can be expressed in terms of the perturbing energy-momentum tensor $T^{\alpha\beta}$; the explicit relations are given by Eqs.~(4.17)--(4.20) of Martel and Poisson. These quantities can therefore be viewed as source terms in the linearized field equations. 

\section{Mukkamala-Pere\~niguez master function}
\label{MP}

In the notation used here, the Mukkamala-Pere\~niguez (MP) master function \cite{mukkamala-pereniguez:24} is defined by
\begin{equation}
\psi_{\rm MP} := -2 r^2 r^a \nabla_a \tilde{K} + (\ell-1)(\ell+2) r \tilde{K} + 2 r r^a r^b \tilde{h}_{ab}.
\label{MP_def1}
\end{equation}
Because it is constructed from gauge-invariant quantities, the MP function is itself gauge invariant. In the standard $(t,r)$ coordinates, the definition becomes
\begin{equation}
\psi_{\rm MP} = -2 r^2 f \partial_r \tilde{K} + (\ell-1)(\ell+2) r \tilde{K} + 2 r f^2 \tilde{h}_{rr},
\label{MP_def2} 
\end{equation}
where $f := 1-2M/r$. 

In their paper, Mukkamala and Pere\~niguez demonstrate that when the perturbing energy-momentum tensor $T^{\alpha\beta}$ vanishes, the Einstein field equations imply that $\psi_{\rm MP}$ satisfies the Regge-Wheeler equation. The generalization to nonvacuum situations is
\begin{equation}
(\Box - V) \psi_{\rm MP} = S,
\label{RW_eq}
\end{equation}
where $\Box := g^{ab} \nabla_a \nabla_b$ is the wave operator in the two-dimensional submanifold of the unperturbed Schwarzschild spacetime,
\begin{equation}
V := \frac{\ell(\ell+1)}{r^2} - \frac{6M}{r^3}
\end{equation}
is the Regge-Wheeler potential (usually associated with odd-parity metric perturbations), and
\begin{align}
S &:= -2 r^2 r^a \nabla_a Q + (\ell^2 + \ell - 4) r Q + \frac{6M}{r} r^a \nabla_a Q^\sharp
- \frac{1}{r} \biggl[ \frac{1}{2} (\ell-1)\ell(\ell+1)(\ell+2) + \frac{12M}{r} - \frac{36M^2}{r^2} \biggr] Q^\sharp
\nonumber \\ & \quad \mbox{} 
+ 2 \biggl[ \ell(\ell+1) - \frac{6M}{r} \biggr] r_a Q^a + 2 r f Q^\flat
\label{source_cov}
\end{align}
is the source term; here $Q := g_{ab} Q^{ab}$.

In the usual $(t,r)$ coordinates, Eq.~(\ref{RW_eq}) becomes
\begin{equation}
\biggl( -\frac{\partial^2}{\partial t^2}
+ f \frac{\partial}{\partial r} f \frac{\partial}{\partial r} - f V \biggr) \psi_{\rm MP} = f S
\label{RW_tr}
\end{equation}
with
\begin{align}
S &= -2 r^2 f \partial_r Q + (\ell^2 + \ell - 4) r Q + \frac{6M}{r} f \partial_r Q^\sharp
- \frac{1}{r} \biggl[ \frac{1}{2} (\ell-1)\ell(\ell+1)(\ell+2) + \frac{12M}{r} - \frac{36M^2}{r^2} \biggr] Q^\sharp
\nonumber \\ & \quad \mbox{} 
+ 2 \biggl[ \ell(\ell+1) - \frac{6M}{r} \biggr] Q^r + 2 r f Q^\flat 
\label{source_tr}
\end{align}
and $Q = -f Q^{tt} + f^{-1} Q^{rr}$.

The easiest way to establish Eq.~(\ref{RW_eq}) is to adopt the Regge-Wheeler gauge in which $j_a = 0 = G$, so that $\tilde{h}_{ab} = h_{ab}$ and $\tilde{K} = K$, and to work in $(t, r)$ coordinates. The first step is to insert Eq.~(\ref{MP_def2}) within the left-hand side of Eq.~(\ref{RW_tr}). The second step is to hunt for the linear superposition of $Q^{ab}$, $Q^a$, $Q^\flat$, and $Q^\sharp$ (as well as their derivatives) that reproduces the left-hand side of the equation; the manipulations make explicit use of Eqs.~(4.13)--(4.16) of Martel and Poisson \cite{martel-poisson:05}, and they eventually return Eq.~(\ref{source_tr}). The third and final step is to make $S$ covariant, as it appears in Eq.~(\ref{source_cov}).

\section{Radiation at future null infinity}
\label{sec:radiation_infinity}

In their Sec.~VI, Martel and Poisson \cite{martel-poisson:05} examine the part of the metric perturbation $p_{\alpha\beta}$ that describes gravitational waves traveling to future null infinity. They work in a radiation gauge in which $t^a p_{ab} = 0 = t^a p_{aB}$, where $t^a$ is the timelike Killing vector of the Schwarzschild spacetime, they adopt coordinates $x^a = (u,r)$, where $u := t - x$ is retarded time (with $x := \int f^{-1}\, dr = r + 2M\ln(r/2M-1)$ denoting the familiar tortoise radius), and they examine the asymptotic behavior of the perturbation ($r \to \infty$ with $u$ fixed), They conclude that the radiation is captured by the components $p_{AB}$ of the metric perturbation, and that these are given asymptotically by
\begin{equation}
p_{AB} \sim r\, \psi_{\rm rad}(u)\, Y_{AB}^{\ell m}, 
\label{radiation_scri}
\end{equation}
where $\psi_{\rm rad}$ is a function of retarded time $u$ that is left undetermined when integrating the vacuum Einstein field equations in a neighborhood of future null infinity. Martel and Poisson go on to show that the flux of gravitational-wave energy that reaches future null infinity is given by
\begin{equation}
\frac{dE}{du} = \frac{1}{64\pi} (\ell-1)\ell(\ell+1)(\ell+2) \biggl| \frac{d \psi_{\rm rad}}{du} \biggr|^2.
\end{equation}
It should be kept in mind that in this equation, $\ell m$ labels are omitted on $\psi_{\rm rad}$, and the right-hand side includes an implicit sum over these labels. A further result established by Martel and Poisson is that
\begin{equation}
\psi_{\rm ZM}(u, r=\infty) = \psi_{\rm rad}(u),
\label{ZM-vs-rad_infinity}
\end{equation}
where the left-hand side denotes the asymptotic limit of the Zerilli-Moncrief master function $\psi_{\rm ZM}(u,r)$, which satisfies the Zerilli equation (instead of the Regge-Wheeler equation).

It is a simple matter to recycle the calculations of Martel and Poisson to determine the relation between $\psi_{\rm rad}$ and the Mukkamala-Pere\~niguez master function. The result is
\begin{equation}
\psi_{\rm MP}(u, r=\infty) = \frac{1}{2} (\ell-1)\ell(\ell+1)(\ell+2)\, \psi_{\rm rad}
+ 6 M \frac{d}{du} \psi_{\rm rad}, 
\end{equation}
and it reveals that the radiation field cannot be expressed algebraically in terms of the MP function. One must instead integrate the differential equation
\begin{equation}
\frac{d}{du} \psi_{\rm rad} + k \psi_{\rm rad} = \frac{1}{6M} \psi_{\rm MP}(u, r=\infty), 
\label{deq_infinity}
\end{equation}
in which the parameter
\begin{equation}
k := \frac{(\ell-1)\ell(\ell+1)(\ell+2)}{12 M}
\label{k_def}
\end{equation}
is recognized as the (purely imaginary) frequency of the algebraically special perturbation of the Schwarzschild spacetime \cite{chandrasekhar:84}. Whether this connection is fortuitous or of deep significance remains to be elucidated. 
 
The solution to Eq.~(\ref{deq_infinity}) is
\begin{equation}
\psi_{\rm rad}(u) = e^{-ku} \biggl[ \psi_{\rm rad}(0)
+ \frac{1}{6M} \int_0^u e^{ku'} \psi_{\rm MP}(u', r=\infty)\, du' \biggr].
\end{equation}
This may be compared with Eq.~(\ref{ZM-vs-rad_infinity}). The comparison reveals that the Zerilli-Moncrief master function is better suited to the description of the radiation field at future null infinity. 

\section{Radiation at the event horizon}
\label{sec:radiation_horizon} 

The radiation that crosses the black-hole horizon is discussed in Sec.~VII of Martel and Poisson \cite{martel-poisson:05}. They again adopt a radiation gauge with $t^a p_{ab} = 0 = t^a p_{aB}$, but now work with coordinates $x^a = (v,r)$, where $v := t + x$ is advanced time. They again show that the radiation field is captured by the angular components of the metric perturbation, this time evaluated at $r=2M$. These are given by 
\begin{equation}
p_{AB} = 2M\, \psi_{\rm rad}(v)\, Y_{AB}^{\ell m}, 
\label{radiation_horizon}
\end{equation}
where $\psi_{\rm rad}$ is a function of $v$ that is left undetermined when integrating the Einstein field equations in a neighborhood of the horizon. The flux of gravitational-wave energy that crosses the horizon is 
\begin{equation}
\frac{dE}{dv} = \frac{1}{64\pi} (\ell-1)\ell(\ell+1)(\ell+2) \biggl| \frac{d \psi_{\rm rad}}{dv} \biggr|^2, 
\end{equation}
where a summation over $\ell$ and $m$ is implied. 

Martel and Poisson show that the radiation field is simply related to the Zerilli-Moncrief master function, 
\begin{equation}
\psi_{\rm ZM}(v, r=2M) = \psi_{\rm rad}(v). 
\label{ZM-vs-rad_horizon}
\end{equation}
The relation with the Mukkamala-Pere\~niguez function, however, is given by 
\begin{equation}
\psi_{\rm MP}(v, r=2M) = \frac{1}{2} (\ell-1)\ell(\ell+1)(\ell+2)\, \psi_{\rm rad}
- 6 M \frac{d}{dv} \psi_{\rm rad}.  
\end{equation}
To express the radiation field in terms of the master function we must therefore integrate the differential equation
\begin{equation}
\frac{d}{dv} \psi_{\rm rad} - k \psi_{\rm rad} = -\frac{1}{6M} \psi_{\rm MP}(v, r=2M), 
\label{deq_horizon}
\end{equation}
in which $k$ is defined by Eq.~(\ref{k_def}). Notice the different signs compared with Eq.~(\ref{deq_infinity}). 

The physically acceptable solution to Eq.~(\ref{deq_horizon}) is
\begin{equation}
\psi_{\rm rad}(v) = \frac{1}{6M} \int_v^\infty e^{-k(v'-v)} \psi_{\rm MP}(v', r=2M)\, dv'.
\label{psi_v} 
\end{equation}
The most general solution to Eq.~(\ref{deq_horizon}) also includes a term $c\, e^{kv}$, with $c$ denoting a constant of integration. This contribution must be eliminated to ensure that $\psi_{\rm rad}$ remains bounded in the remote future. Equation (\ref{psi_v}) then enforces the teleological boundary condition that $\psi_{\rm rad}(v=\infty) = 0$. A comparison with Eq.~(\ref{ZM-vs-rad_horizon}) reveals that the Zerilli-Moncrief master function is better suited to the description of the radiation field at the black-hole horizon.  

\section{Metric reconstruction}
\label{sec:reconstruction} 

In this section I consider the task of reconstructing the metric perturbation from the Mukkamala-Pere\~niguez master function. I adopt the Regge-Wheeler gauge and work in the coordinates $x^a = (t,r)$. Martel previously established \cite{martel:04} that in these gauge and coordinates, $h_{ab}$ and $K$ can be obtained from the Zerilli-Moncrief function $\psi_{\rm ZM}$ and the sources $Q^{ab}$, $Q^a$, $Q^\sharp$, and $Q^\flat$. The reconstruction is entirely explicit, in the sense that $h_{ab}$ and $K$ are expressed directly in terms of $\psi_{\rm ZM}$ and its derivatives, and in terms of the sources and their derivatives. 

The situation appears to be more involved in the case of the MP function. After an exhaustive search, I am left unable to state an explicit expression for $K$ in terms of $\psi_{\rm MP}$ and sources. The simplest option appears to be the differential equation
\begin{equation}
f \partial_r K - k K = -\frac{1}{6M} \biggl[ f \partial_r \psi_{\rm MP}
+ \frac{\ell(\ell+1)}{2r} \psi_{\rm MP} - 2r^2 f^2 Q^{tt} \biggr],
\label{K_vs_MP} 
\end{equation}
with $k$ still defined by Eq.~(\ref{k_def}). Once $K$ is obtained by integrating Eq.~(\ref{K_vs_MP}), $h_{ab}$ is recovered explicitly:
\begin{subequations}
\begin{align}
h_{rr} &= \frac{1}{2 r f^2} \Bigl[ \psi_{\rm MP} + 2 r^2 f \partial_r K
- (\ell-1)(\ell+2) r K \Bigr], \\
h_{tr} &= -\frac{1}{\ell(\ell+1) f} \Bigl\{ \partial_t \psi_{\rm MP}
- r \bigl[ \ell(\ell+1) - 6M/r \bigr] \partial_t K + 2r^2 f Q^{tr} \Bigr\}, \\
h_{tt} &= f^2 h_{rr} + f Q^\sharp.
\end{align}
\end{subequations}
Of course, my failure to find an explicit expression for $K$ does not constitute a proof that such an expression does not exist; others might have better luck or more stamina. There might also be other choices of gauge for which the metric reconstruction can be achieved explicitly. On the basis of the current evidence, however, I tentatively conclude that the Zerilli-Moncrief master function is better suited to the task of metric reconstruction.

The physically acceptable solution to Eq.~(\ref{K_vs_MP}) is
\begin{equation}
K(t,x) = -\int_x^\infty e^{-k(x'-x)} S_K(t, x')\, dx',
\end{equation}
where $x := \int f^{-1}\, dr = r + 2M\ln(r/2M - 1)$ and $S_K$ stands for the right-hand side of Eq.~(\ref{K_vs_MP}), re-expressed in terms of the tortoise variable $x'$. The general solution to the differential equation also includes a term $c\, e^{kx}$, with $c$ denoting a constant of integration; this was eliminated to ensure that $K$ stays bounded at $r = \infty$. 

\begin{acknowledgments} 
Exchanges with David Pere\~niguez were greatly appreciated. This work was supported by the Natural Sciences and Engineering Research Council of Canada.  
\end{acknowledgments} 

\bibliography{mp}
\end{document}